\documentclass{article}

\usepackage[T1]{fontenc}
\usepackage[utf8]{inputenc}
\usepackage[]{ismir} % Remove the "submission" option for camera-ready version
\usepackage{caption}
\usepackage{amsmath,cite,url}
\usepackage{graphicx}
\usepackage{color}

\usepackage{booktabs}
\usepackage{tabularx}
\usepackage{multirow}
\usepackage[bookmarks=false,hidelinks]{hyperref}
\usepackage{acro}
\usepackage{enumitem}
\usepackage{siunitx}
\usepackage{array}
\usepackage{xcolor}
\usepackage{float}
\usepackage[hyphens]{xurl}

\DeclareAcronym{vi}{short=VI, long=version identification}
\DeclareAcronym{ti}{short=TI, long=track identification}
\DeclareAcronym{cqt}{short=CQT, long=Constant-Q transform}
\DeclareAcronym{qbe}{short=QbE, long=query-by-example}
\DeclareAcronym{ann}{short=ANN, long=approximate nearest neighbor}
\DeclareAcronym{ir}{short=IR, long=impulse response}
\DeclareAcronym{ivf}{short=IVF, long=inverted file list}
\DeclareAcronym{map}{short=MAP, long=mean average precision}
\DeclareAcronym{nar}{short=NAR, long=normalized average rank}
\DeclareAcronym{hr}{short=HR, long=hit rate}
\DeclareAcronym{wp}{short=w.p., long=with probability}
\DeclareAcronym{er}{short=ER, long=exhaustive retrieval}
\DeclareAcronym{amp}{short=AMP, long=automatic mixed precision}

\newcommand{\hratk}[1]{\ac{hr}@#1}
\newcommand{\mapat}[1]{MAP@#1}
\newcommand{\narat}[1]{NAR@#1}
\newcommand{\mandeg}{manipulated\,\&\,degraded}

\newcommand{\discogsvi}{Discogs-VI}
\newcommand{\shs}{SHS100K}

\newcommand{\clews}{CLEWS}
\newcommand{\grafprint}{GraFPrint}
\newcommand{\peaknet}{PeakNetFP}
\newcommand{\clapp}{CLAP}
\newcommand{\nmfp}{NMFP}

\newcommand{\bytecover}{ByteCover2}
\newcommand{\bytecoverr}{ByteCover3}
\newcommand{\coverhunter}{CoverHunter}
\newcommand{\ours}{Fish}

\newcolumntype{R}{>{\hspace*{4pt}}r}
\newcolumntype{G}{>{\hspace*{8pt}}r}
\newcolumntype{C}{@{\,\textcolor{gray}{\footnotesize$\pm$}\,}>{\color{gray}\footnotesize}l}
\newcommand{\B}[1]{\text{\textbf{#1}}}
\newcommand{\U}[1]{\text{\underline{#1}}}

\title{Unified Music Identification for Tracks and Versions}

\multauthor
  {R.\ Oguz Araz$^1$ \quad Joan Serr\`a$^2$ \quad Yuki Mitsufuji$^{2,3}$ \quad Xavier Serra$^1$ \quad Dmitry Bogdanov$^1$\vspace{2mm}}{$^1$ Music Technology Group, Universitat Pompeu Fabra, Barcelona\\
  $^2$ Sony AI \quad $^3$ Sony Group Corporation\\
  {\tt\small recepoguz.araz@upf.edu}
  }

\def\authorname{R.\ O.\ Araz, J.\ Serr\`a, Y.\ Mitsufuji, X.\ Serra, and D.\ Bogdanov}

\begin{document}

\maketitle

\begin{abstract}
Given a music database, \ac{ti} retrieves the exact track matching an audio excerpt, whereas \ac{vi} retrieves its musical versions.
Traditionally, the two tasks have been addressed separately.
However, as \textit{every track is its own closest version}, we investigate whether \ac{vi} can subsume \ac{ti}.
This requires \ac{vi} systems to be robust to both signal manipulation and audio degradation.
We therefore propose a unified benchmark that evaluates accuracy and robustness on each task.
Comparing seven existing models on this benchmark, we show that none of them are both accurate and robust on both tasks.
We then train a baseline model targeting both tasks and show that a unified system is possible with 10\,s \ac{ti} queries.
Lastly, we characterize the two retrieval constraints that limit our model's \ac{ti} performance.
We envision extending this unification to other music identification tasks.
\end{abstract}

\acresetall

\section{Introduction}\label{sec:intro}
% the problem and why it matters
In a database of music tracks, \ac{ti} searches for the exact track matching an audio excerpt~\cite{haitsma_robust_2001}, whereas \ac{vi} searches for its musical versions (e.g., covers, remixes, live takes)~\cite{yesiler_audio-based_2021}.
\ac{ti} and \ac{vi} have traditionally existed as separate tasks, although they are related as \textit{every track is its own closest version}.
Therefore, we ask: ``Can \ac{vi} subsume \ac{ti}?''
A unified system could provide identity information on multiple levels, supporting music discovery and rights management.

% the current state and its limitations
Beyond accuracy, \ac{ti} requires robustness to both signal manipulation (e.g., pitch shifting, time stretching) and audio degradation (e.g., room reverberation, background noise)~\cite{cano_review_2002}.
However, most neural \ac{ti} models prioritize robustness to degradation over signal manipulation.
Conversely, \ac{vi} models are trained to be robust to musical variations across versions (e.g., tempo change, key transposition), but rarely to degradation.
To the best of our knowledge, there is no systematic robustness study of \ac{vi} models.

% our approach
To fill these gaps, we propose a unified benchmark that evaluates accuracy and robustness for both tasks over three databases: one shared by both tasks and one specific to each task.
We then compare seven existing models on this benchmark and show that none of them are both accurate and robust on both tasks.
Finally, we train a model from scratch targeting both tasks.
To build its training data, we introduce a method for locating audio segments that match across versions.

% contributions
Our contributions are as follows:
We conduct the largest study to date of embedding-based \ac{ti} and the first systematic robustness study of \ac{vi} models.
We also release timestamps at 1\,s resolution for 97\,M pairs of 20\,s audio segments that contain different renditions of the same musical phrase, grouped into 644\,k \emph{segment cliques}, obtained from 317\,k tracks across 74\,k track cliques.
Finally, we present the first baseline model that targets both tasks.
Code and trained model weights\footnote{\url{https://github.com/raraz15/unified-track-and-version-id}} are shared, together with the metadata\footnote{\url{https://mtg.github.io/discogs-vi-dataset/}}.

\begin{figure*}[t]
    \centering
    \captionsetup{skip=0pt}
    \includegraphics[width=\textwidth]{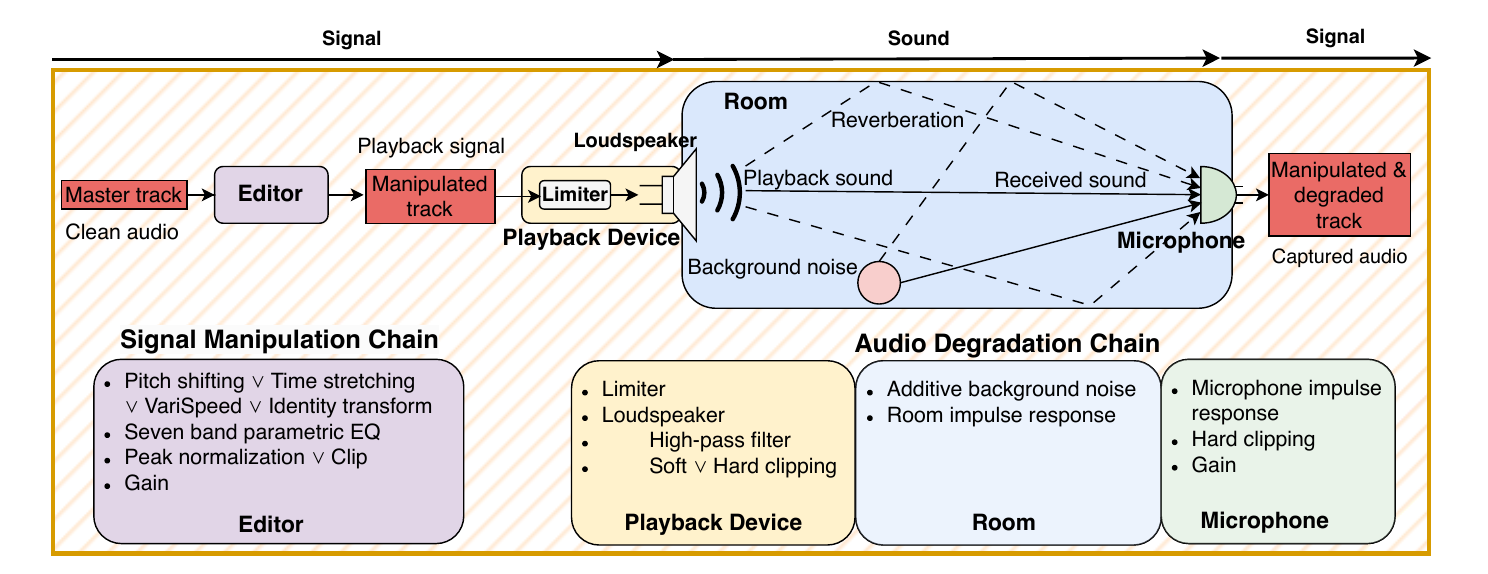}
    \caption{
    Music editing-playback-capture chain. Signals highlighted in red are used as queries during evaluation.
    }
    \label{fig:signal}
\end{figure*}

\section{Related Work}\label{sec:sota}
\ac{ti} and \ac{vi} models differ along three axes central to our study: input duration, training supervision, and robustness scope.
\ac{ti} models embed short audio segments (typically 1\,s) into low-dimensional vectors (typically 128~dimensions) that preserve the acoustic detail characterizing a unique recording~\cite{chang_neural_2021,araz_enhancing_2025,bhattacharjee_grafprint_2025,cortes-sebastia_peaknetfp_2025}.
Most \ac{ti} models are trained with self-supervised learning, where positive pairs are generated by altering a segment, targeting only robustness to audio degradation~\cite{chang_neural_2021,araz_enhancing_2025,bhattacharjee_grafprint_2025}. 
Only a few studies address signal manipulation~\cite{cortes-sebastia_peaknetfp_2025,singh_robust_2025}.

In contrast, \ac{vi} models suppress acoustic detail to match versions~\cite{du_bytecover3_2023,liu_coverhunter_2023,serra_supervised_2025}. They support variable-length inputs and embed longer segments (typically 20\,s) into higher-dimensional vectors (typically 1024~dimensions).
\bytecoverr{}~\cite{du_bytecover3_2023} learns segment representations by fine-tuning \bytecover{}~\cite{du_bytecover2_2022}, whereas \clews{}~\cite{serra_supervised_2025} trains directly on segments from scratch.
Both rely on weak supervision, since version labels do not indicate which segments match across versions.
\coverhunter{}~\cite{liu_coverhunter_2023} instead first trains a classification model and then uses it to locate similar segment pairs for a second, fully supervised training stage.
\ac{vi} models are generally trained with signal manipulation, but only \coverhunter{} includes audio degradation, limited to background noise.

\section{Unified Benchmark}\label{sec:eval}
Our benchmark evaluates accuracy and robustness for both \ac{ti} and \ac{vi} using three databases: a shared database is queried for both tasks while one external database per task is included for additional reference. 
We also propose a comprehensive music degradation and manipulation chain.

\subsection{Music Editing-Playback-Capture Chain}\label{sec:eval:signal}
\figref{fig:signal} depicts the signal chain encountered in various music identification scenarios.
The chain begins with possible \emph{signal manipulation}: a master track being edited for creative purposes or to evade copyright detection.
In many applications, the playback signal is not available, but the playback sound can be recorded through a microphone.
Compared to the playback signal, the microphone output is degraded by the loudspeaker, the acoustic environment, and the microphone itself.
We refer to this combined effect as \emph{audio degradation}.
Following this signal chain, our benchmark evaluates \ac{vi} and \ac{ti} performance under three query conditions: clean (i.e., taken from the master track), manipulated, and \mandeg{}.
We simulate the chain using \texttt{audiomentations}\footnote{\url{https://iver56.github.io/audiomentations/}}, modeling the linear components and, where feasible, the nonlinearities.
We detail the individual components of each stage following the application order, where parameters are sampled uniformly from the given ranges.

\vspace{0.1cm}\noindent
\textbf{Signal manipulation chain:}
Exactly one of the following transforms is applied with equal probability: pitch shifting by $p\!\in\![-12,\,12]$ semitones, time stretching by $t\!\in\![0.5,\,2.0]$, playback speed change (VariSpeed, i.e., rate change that co-varies time\,\&\,pitch) by $s\!\in\![0.5,\,2.0]$, or the identity transform.
Then, \ac{wp}~$0.5$, a seven-band parametric EQ with moving center frequencies is applied, where each band's gain is sampled from $[-12,\,12]\,\mathrm{dB}$.
If a signal exceeds $0\,\mathrm{dBFS}$, we peak-normalize or clip it, each \ac{wp}~$0.5$.
Finally, \ac{wp}~$0.5$, we apply a global gain $g\!\in\![-6,\,0]\,\mathrm{dB}$.

\vspace{0.1cm}\noindent
\textbf{Audio degradation chain:}
We model the playback device as a loudspeaker preceded by a limiter (encountered in music venues for device protection or loudness control).
The limiter is applied \ac{wp}~$0.5$ and has a threshold $T\!\in\![-12,\,-1]\,\mathrm{dBFS}$, attack time $t_a\!\in\![0.5,\,25]\,\mathrm{ms}$, and release time $t_r\!\in\![50,\,100]\,\mathrm{ms}$.
Loudspeakers introduce hard-to-model nonlinearities; hence, as a simple proxy, \ac{wp}~$0.5$, we apply a high-pass filter with a cutoff in $[20,\,150]\,\mathrm{Hz}$ (following~\cite{akesbi_audio_2022}).
Then, \ac{wp}~$0.5$, we add loudspeaker nonlinearity via soft-clipping (TanH function) or hard-clipping (randomly clipping samples above the $x\,\in\,[90,\,100]$ percentile), chosen with equal probability.
We model the acoustic environment with background noise and room reverberation.
We both add background noise at an $\mathrm{SNR}\!\in\![-3,\,15]\,\mathrm{dB}$ (repeating the noise if it is shorter than the music) and convolve the mixture with a room \ac{ir}.
Finally, we model the microphone with a microphone \ac{ir}, followed by random hard-clipping (same parameters as before) \ac{wp}~$0.5$, and an additional global gain from $[-6,\,0]\,\mathrm{dB}$ applied \ac{wp}~$0.5$.

\subsection{Audio Data}\label{sec:eval:data}
We downsample all audio files to 16\,kHz and downmix to mono.
For degradation, we utilize the proposed test sets of \nmfp{}~\cite{araz_enhancing_2025}: background noise, room \ac{ir}, and microphone \ac{ir} recordings.
For music audio, we create three separate test databases using the \discogsvi{}~\cite{araz_discogs-vi_2024}, \shs{}~\cite{xu_key-invariant_2018}, and \nmfp{} test sets.
The music of \nmfp{} is taken from the FMA~\cite{defferrard_fma_2017} dataset, which consists mainly of non-professional recordings, while \discogsvi{} exclusively contains official YouTube uploads, and \shs{} combines both types.
\discogsvi{} is used for evaluating both \ac{ti} and \ac{vi}, \shs{} is used only for \ac{vi}, and \nmfp{} is used only for \ac{ti}.
\discogsvi{} uniquely supports both \ac{ti} and \ac{vi} evaluations on a shared database, central to our unification question.
Moreover, \discogsvi{} is a challenging dataset for \ac{ti} because musical versions act as hard distractors.
The databases created from \discogsvi{}, \shs{}, and \nmfp{} contain about 116\,k, 8\,k, and 95\,k~full tracks, respectively.

\subsection{Query Types}\label{sec:eval:query}
Using the signal chain of Section~\ref{sec:eval:signal}, we create three query types from each reference track: (i) clean, (ii) manipulated, and (iii) \mandeg{}.
In \ac{ti}, a query should be identified independently of where it lies in the reference track, so previous work randomly samples segments as queries~\cite{chang_neural_2021,araz_enhancing_2025}.
Following this practice, we randomly sample 10\,s of audio per track, aligning start times across query types by accounting for possible time-stretch.
Our 10\,s duration reflects \ac{ti} applications such as consumer music recognition and broadcast monitoring, where several seconds of query audio are typically available.

In \ac{vi}, however, a single random segment is rarely sufficient for identification, since not all segment pairs across versions contain renditions of the same phrase (e.g., a chorus and bridge pair).
Therefore, in the absence of phrase-level annotations, we query entire tracks for \ac{vi}, following~\cite{serra_supervised_2025}.
This yields about 372\,k full-track \ac{vi} queries (124\,k tracks $\times$ 3 query types) and 633\,k segment \ac{ti} queries (211\,k segments $\times$ 3 query types), totaling over 1\,M queries.
For \ac{ti} in particular, this contrasts sharply with prior work, which typically queries a few hundred segments in total~\cite{chang_neural_2021,cortes-sebastia_peaknetfp_2025,su_amg-embedding_2024,bhattacharjee_grafprint_2025,chen_variable-length_2026}.

\subsection{Compared Models and Input Segmentation}\label{sec:eval:models}
For \ac{ti}, we consider \nmfp{}~\cite{araz_enhancing_2025}, \grafprint{}~\cite{bhattacharjee_grafprint_2025}, and \peaknet{}~\cite{cortes-sebastia_peaknetfp_2025}.
For \ac{vi}, we consider \clews{}~\cite{serra_supervised_2025}, \bytecoverr{}~\cite{du_bytecover3_2023}, and \coverhunter{}~\cite{liu_coverhunter_2023}.
We also benchmark \clapp{}~\cite{wu_large-scale_2023}, a language–audio representation model.
We use official implementations except \bytecoverr{}, for which we use the \bytecover{}~\cite{du_bytecover2_2022} implementation from~\cite{serra_supervised_2025}.

We segment full tracks with each model's training context duration: 1\,s for \nmfp{}, \grafprint{}, and \peaknet{}; 20\,s for \clews{} and \bytecover{}; 45\,s for \coverhunter{}; and 10\,s for \clapp{}. 
The hop duration is fixed at 0.5\,s for \ac{ti} models and 5\,s for the rest, guaranteeing at least 50\% overlap between consecutive segments; the 5\,s hop is additionally bounded from below by our computational budget of 4\,h per query type per \ac{vi} evaluation.
Meanwhile, the \ac{ti} queries are 10\,s segments (Section~\ref{sec:eval:query}), which variable-length models such as \clews{} embed into a single vector, whereas models such as \nmfp{} embed into 19 vectors.

\subsection{Retrieval}\label{sec:eval:retrieval}

We perform retrieval using segment-level embeddings instead of aggregating into track-level embeddings by temporal averaging~\cite{affolter_scalable_2026} or learned aggregators~\cite{su_amg-embedding_2024}.
Exhaustive retrieval with segments~\cite{serra_supervised_2025,mancini_leveraging_2026} requires substantial GPU memory and long runtimes at the scale of \discogsvi{}.
Therefore, following common practice in the \ac{ti} literature, we implement \ac{ann} retrieval using NVIDIA \texttt{cuVS}~\cite{feher_accelerated_2023}.
We utilize \ac{ivf}-Flat indices, which partition the embedding space into $n_{\text{lists}}$ clusters.
At search time, for each query vector, the algorithm probes the $n_{\text{probes}}$ closest clusters and returns the top-$k$ most similar segments.
We set $n_{\text{lists}}$ to the square root of the database size, following~\cite{feher_accelerated_2023}.
Achieving sufficient recall in \ac{vi} requires more probing because there can be multiple relevant tracks per query.
Therefore, we set $n_{\text{probes}}$ to 1\% and 3\% of $n_{\text{lists}}$, and $k$=1,024 and $k$=10,240 for \ac{ti} and \ac{vi}, respectively.
For a query sequence of length $T$, the index yields up to $T{\cdot}k$ candidate segments with the corresponding distances.
We keep the best-scoring segment per track for ranking.
We measure \ac{ti} performance using \hratk{1} and \hratk{10}~\cite{chang_neural_2021}, and \ac{vi} performance using \mapat{N} and \narat{N} (normalized average rank~\cite{serra_supervised_2025}) with N=10\,k.
% \footnote{\url{https://github.com/rapidsai/cuvs}}%
% \ac{map} and \ac{nar} metrics~\cite{serra_supervised_2025}.
% Since \ac{ann} returns only a subset, we use the cut-off variants

\section{A Baseline for Unified Identification}\label{sec:baseline}
Following our research premise, we train a \ac{vi} model that targets both tasks.
\clews{} and \bytecoverr{} learn to match the musical phrases that are shared between versions without explicit supervision.
Instead, inspired by \coverhunter{}'s second stage, we adopt fully supervised learning.
A novelty of our work is the training data: cliques of 20\,s audio segments that contain different versions of the same musical \emph{phrase}.
Drawing an analogy to \emph{learning to fish} versus \emph{being given fish}, we call our model \ours{}.

\subsection{Segment Cliques}\label{sec:baseline:data}

\begin{figure}[tb]
    \centering
    \captionsetup{skip=0pt}
    \includegraphics[width=\columnwidth]{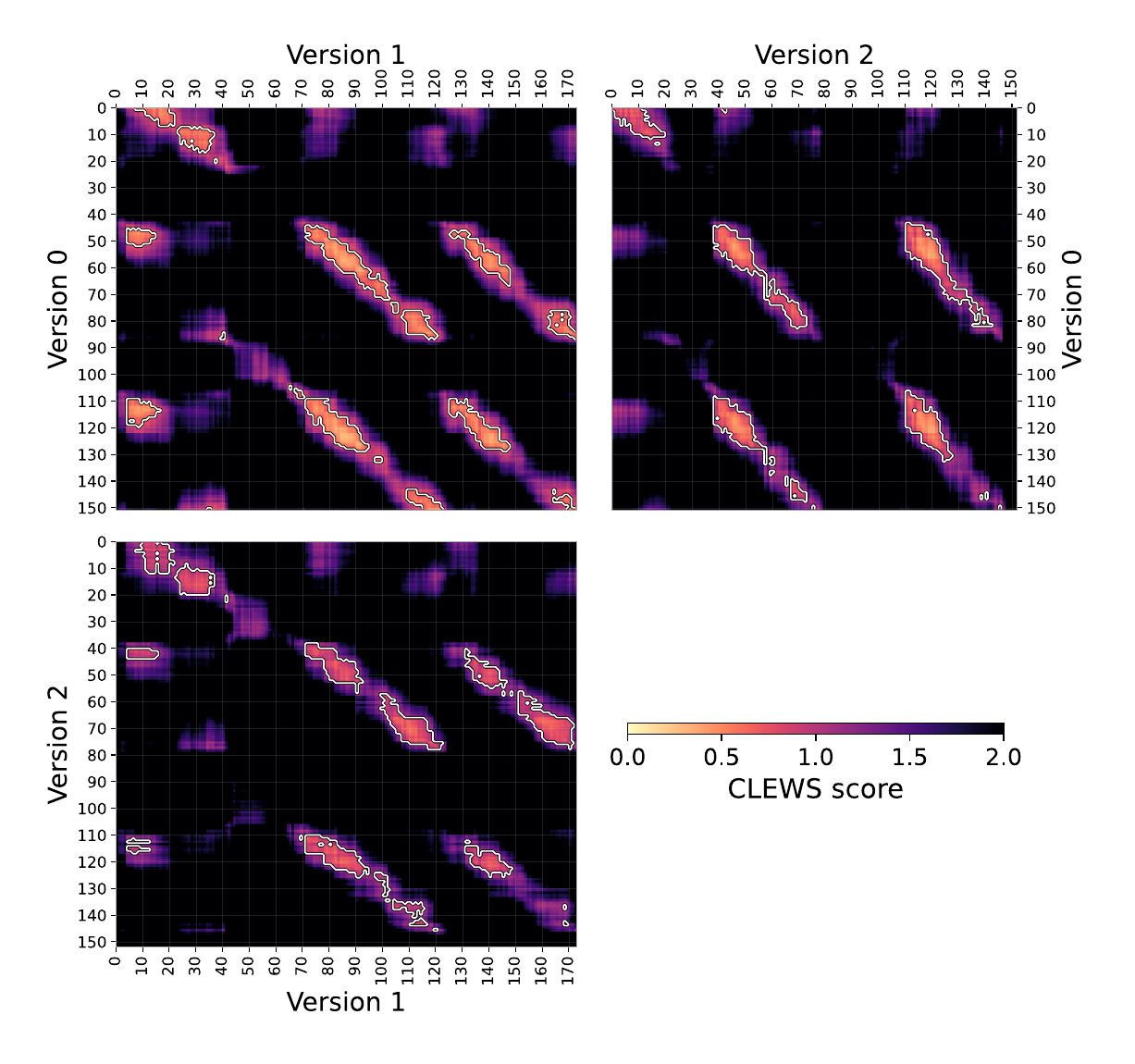}
    \caption{
    Pairwise segment-level distance matrices of 3 musical versions, computed on audio segments of 20\,s with 1\,s hop, obtained with \clews{}.
    Similar regions located by our method are outlined by the white contour lines.
    }
    \label{fig:cliques}
\end{figure}

To obtain segment cliques, we first cut the \discogsvi{} train set tracks into 20\,s segments with a 1\,s hop and embed each segment with the pre-trained \clews{} model.
For every version pair, we compute a segment-level distance matrix using the native \clews{} distance, where lower values indicate higher similarity.
We discard silent segments (via an energy threshold) and segments classified as non-music by~\cite{kong_panns_2020} (AudioSet classes mapped following~\cite{araz_evaluation_2024}) by setting their distances to infinity.
The resulting matrices resemble the topography of a river basin (a heightfield over the x-y plane), examples of which are shown in \figref{fig:cliques}.

These distance matrices are noisy: (1) not all segment pairs contain versions of the same musical phrase (false positives), and (2) \clews{} does not assign small distances to all true positives (false negatives).
Therefore, a threshold is needed to reject false positives. 
% without discarding true positives.
A static distance threshold across version pairs (as used by \coverhunter{}) may admit false positives for similar versions (e.g., same-genre versions) and reject true positives for hard versions (e.g., cross-genre versions).
Instead, we set the threshold dynamically as the 5\textsuperscript{th} percentile distance of each matrix.

The thresholded matrices have regions with high similarity, whose boundaries we find using the \texttt{flood-fill} algorithm\footnote{\url{https://scikit-image.org/docs/stable/api/skimage.morphology.html\#skimage.morphology.flood}}.
From a given point, it finds the connected neighbors in the x-y plane below a height threshold.
Each local minimum is called a \emph{hole}, and the connected sub-threshold neighbors, a \emph{basin}.
We find all basins iteratively, starting from the lowest hole.
For training, we use only the hole of each basin as its representative pair, as basins contain numerous redundant pairs clustered around this high-confidence pair.
% It also removes many true-positives due to overflooding
Finally, we group the holes that overlap by 19\,s into \emph{segment} cliques.
This grouping de-duplicates segments, balancing their representation during training.
It also enforces similarity transitivity between segments, restoring hard pairs that the distance threshold discards.
% We refer to the code for the full details of the algorithm.

\begin{table*}[t]
\centering
\captionsetup{skip=5pt}
\setlength{\tabcolsep}{11pt}
\resizebox{\textwidth}{!}{
\begin{tabular}{cl@{\hspace{3em}} rC rC GC rC rC GC}
\toprule
    \multirow{2}[+2]{*}{\textbf{Metric}} & \multirow{2}[+2]{*}{\textbf{Model}}
    & \multicolumn{6}{c}{\textbf{\nmfp{} Test}} & \multicolumn{6}{c}{\textbf{\discogsvi{} Test}} \\
\cmidrule(lr){3-8}\cmidrule(lr){9-14}
     & & \multicolumn{2}{c}{\textbf{Clean}} & \multicolumn{2}{c}{\textbf{Manip.}} & \multicolumn{2}{c}{\textbf{Manip. \& Deg.}}
       & \multicolumn{2}{c}{\textbf{Clean}} & \multicolumn{2}{c}{\textbf{Manip.}} & \multicolumn{2}{c}{\textbf{Manip. \& Deg.}} \\
\midrule
    \multirow{8}{*}{\rotatebox[origin=c]{90}{\textbf{\hratk{1} ($\uparrow$)}}}
         & \grafprint{}   & 83.7 & 0.2 & 41.3 & 0.3 & 11.7 & 0.2 & 71.0 & 0.3 & 36.3 & 0.3 & 12.0 & 0.2 \\
         & \peaknet{}     & \U{94.8} & 0.1 & 31.2 & 0.3 & 2.4  & 0.1 & \U{95.9} & 0.1 & 32.1 & 0.3 & 2.7  & 0.1 \\
         & \nmfp{}        & \B{96.9} & 0.1 & 48.7 & 0.3 & \B{45.5} & 0.3 & \B{97.4} & 0.1 & 48.8 & 0.3 & \U{47.2} & 0.3 \\
         & \clapp{}       & 89.9 & 0.2 & 43.2 & 0.3 & 3.2  & 0.1 & 84.5 & 0.2 & 32.1 & 0.3 & 1.9  & 0.1 \\
         % & \alain{}       &   &  &  &  &  &  & 96.0 & 0.1 & 63.6 & 0.3 & 21.8 & 0.2 \\
         & \coverhunter{} & 10.8 & 0.2 & 5.0  & 0.1 & 1.5  & 0.1 & 10.8 & 0.2 & 6.0  & 0.1 & 1.5  & 0.1 \\
         & \bytecover{}   & 67.1 & 0.3 & 43.2 & 0.3 & 8.6  & 0.2 & 70.4 & 0.3 & 52.8 & 0.3 & 16.7 & 0.2 \\
         & \clews{}       & 86.4 & 0.2 & \U{50.1} & 0.3 & 20.8 & 0.3 & 78.0 & 0.2 & \U{55.2} & 0.3 & 31.1 & 0.3 \\
         % & \ours{}        & 86.2 & 0.2 & \B{69.8} & 0.3 & \U{39.5} & 0.3 & 83.4 & 0.2 & \B{74.9} & 0.3 & \B{54.0} & 0.3 \\
         % & \ours{}        & 87.5 & 0.2 & \B{69.8} & 0.3 & \U{41.0} & 0.3 & 85.5 & 0.2 & \B{75.9} & 0.2 & \B{55.6} & 0.3 \\
         & \ours{} & 86.9 & 0.2 & \B{69.7} & 0.3 & \U{41.6} & 0.3 & 83.7 & 0.2 & \B{74.4} & 0.3 & \B{53.6} & 0.3 \\
\midrule
    \multirow{8}{*}{\rotatebox[origin=c]{90}{\textbf{\hratk{10} ($\uparrow$)}}}
         & \grafprint{}   & 92.6 & 0.2 & 49.8 & 0.3 & 19.4 & 0.3 & 83.5 & 0.2 & 45.1 & 0.3 & 19.9 & 0.2 \\
         & \peaknet{}     & \U{99.3} & 0.1 & 38.4 & 0.3 & 6.3  & 0.2 & \U{99.5} & 0.0 & 39.7 & 0.3 & 6.9  & 0.1 \\
         & \nmfp{}        & \B{99.6} & 0.0 & 50.6 & 0.3 & \U{48.5} & 0.3 & \B{99.7} & 0.0 & 50.1 & 0.3 & 49.5 & 0.3 \\
         & \clapp{}       & 96.9 & 0.1 & 58.9 & 0.3 & 7.4  & 0.2 & 94.5 & 0.1 & 47.7 & 0.3 & 5.1  & 0.1 \\
         % & \alain{}       &   &  &  &  &  &  & 99.9 & 0.0 & 69.6 & 0.3 & 27.1 & 0.3 \\
         & \coverhunter{} & 19.9 & 0.2 & 12.5 & 0.2 & 3.9  & 0.1 & 19.9 & 0.2 & 12.5 & 0.2 & 3.9  & 0.1 \\
         & \bytecover{}   & 74.0 & 0.3 & 53.0 & 0.3 & 13.0 & 0.2 & 77.2 & 0.2 & 63.5 & 0.3 & 24.4 & 0.2 \\
         & \clews{}       & 93.5 & 0.2 & \U{62.5} & 0.3 & 30.1 & 0.3 & 92.8 & 0.1 & \U{77.1} & 0.2 & \U{49.6} & 0.3 \\
         % & \ours{}        & 92.9 & 0.2 & \B{80.2} & 0.3 & \B{50.3} & 0.3 & 92.8 & 0.1 & \B{87.6} & 0.3 & \B{69.6} & 0.3 \\
         % & \ours{} & 93.8 & 0.2 & \B{80.3} & 0.3 & \B{52.1} & 0.3 & 94.1 & 0.1 & \B{88.5} & 0.2 & \B{71.3} & 0.3 \\
         & \ours{} & 93.2 & 0.2 & \B{80.1} & 0.3 & \B{52.7} & 0.3 & 93.1 & 0.1 & \B{87.3} & 0.2 & \B{69.4} & 0.3 \\
\bottomrule
\end{tabular}
}
\caption{
\ac{ti} benchmark results. 
The $\pm$ symbol denotes a 95\% confidence interval. 
}
\label{tab:track}
\end{table*}

\begin{table*}[th]
\centering
\captionsetup{skip=5pt}
\setlength{\tabcolsep}{8pt}
\resizebox{\textwidth}{!}{
\begin{tabular}{cl@{\hspace{3em}} rC rC RC rC rC RC}
\toprule
    \multirow{2}[+2]{*}{\textbf{Metric}} & \multirow{2}[+2]{*}{\textbf{Model}} & \multicolumn{6}{c}{\textbf{\discogsvi{} Test}} & \multicolumn{6}{c}{\textbf{\shs{} Test}} \\
\cmidrule(lr){3-8}\cmidrule(lr){9-14}
    & & \multicolumn{2}{c}{\textbf{Clean}} & \multicolumn{2}{c}{\textbf{Manip.}} & \multicolumn{2}{c}{\textbf{Manip. \& Deg.}} & \multicolumn{2}{c}{\textbf{Clean}} & \multicolumn{2}{c}{\textbf{Manip.}} & \multicolumn{2}{c}{\textbf{Manip. \& Deg.}} \\
\midrule
    \multirow{7}{*}{\rotatebox[origin=c]{90}{\textbf{\mapat{10\,k} ($\uparrow$)}}}
         & \grafprint{}  & 0.014 & 0.001 & 0.007 & 0.000 & 0.005 & 0.000 & 0.010 & 0.001 & 0.007 & 0.000 & 0.007 & 0.000 \\
         & \nmfp{}       & 0.010 & 0.000 & 0.005 & 0.000 & 0.009 & 0.000 & 0.010 & 0.001 & 0.006 & 0.000 & 0.008 & 0.000 \\
         & \clapp{}      & 0.019 & 0.001 & 0.010 & 0.000 & 0.002 & 0.000 & 0.027 & 0.001 & 0.021 & 0.001 & 0.009 & 0.001 \\
         % & \alain{}       & 0.054 & 0.001 & 0.029 & 0.001 & 0.011 & 0.000 \\
         & \coverhunter{}& 0.031 & 0.001 & 0.019 & 0.001 & 0.006 & 0.000 & 0.129 & 0.006 & 0.086 & 0.005 & 0.036 & 0.003 \\
         & \bytecover{}  & 0.273 & 0.002 & 0.208 & 0.002 & 0.164 & 0.002 & 0.469 & 0.008 & 0.357 & 0.008 & 0.276 & 0.007 \\
         & \clews{}      & \U{0.628} & 0.002 & \U{0.604} & 0.002 & \U{0.523} & 0.002 & \U{0.805} & 0.006 & \U{0.751} & 0.007 & \U{0.668} & 0.008 \\
         % & \ours{} 1 gpu no noise      & \B{0.654} & 0.002 & \B{0.616} & 0.002 & \B{0.573} & 0.002 & \U{0.792} & 0.007 & \B{0.755} & 0.007 & \B{0.727} & 0.007 \\
         % & \ours{}         & \B{0.677} & 0.002 &\B{ 0.641} & 0.002 & \B{0.599} & 0.002 & 0.800 & 0.007 & \B{0.764} & 0.007 & \B{0.737} & 0.007 \\
         & \ours{} & \B{0.687} & 0.002 & \B{0.654} & 0.002 & \B{0.612} & 0.002 & \B{0.806} & 0.007 & \B{0.772} & 0.007 & \B{0.748} & 0.007 \\
\midrule
    \multirow{7}{*}{\rotatebox[origin=c]{90}{\textbf{\narat{10\,k} ($\downarrow$)}}}
         & \grafprint{}  & 82.0 & 0.1 & 84.4 & 0.1 & 84.0 & 0.1 & 35.1 & 0.4 & 38.4 & 0.4 & 39.0 & 0.4 \\
         & \nmfp{}       & 84.0 & 0.1 & 86.9 & 0.1 & 85.2 & 0.1 & 38.7 & 0.4 & 42.7 & 0.4 & 41.7 & 0.4 \\
         & \clapp{}      & 66.3 & 0.2 & 69.7 & 0.2 & 78.4 & 0.1 & 26.4 & 0.5 & 27.6 & 0.4 & 37.0 & 0.5 \\
         % & \alain{}       & 60.2 & 0.2 & 67.6 & 0.2 & 77.7 & 0.1 & \\
         & \coverhunter{}& 65.9 & 0.2 & 70.1 & 0.2 & 78.9 & 0.2 & 15.5 & 0.3 & 19.1 & 0.3 & 26.1 & 0.4 \\
         & \bytecover{}  & 28.3 & 0.2 & 32.7 & 0.2 & 38.5 & 0.2 & 6.5  & 0.2 & 8.7  & 0.3 & 12.1  & 0.3 \\
         & \clews{}      & \U{8.2} & 0.1 & \U{8.6} & 0.1 & \U{11.3} & 0.1 & \B{1.6} & 0.1 & \B{1.9} & 0.2 & \U{2.9} & 0.2 \\
         % & \ours{} 1 gpu no noise      & \B{8.2} & 0.1 & \U{8.7} & 0.1 & \B{9.9} & 0.1 & \U{1.7} & 0.2 & \U{2.0} & 0.2 & \B{2.2} & 0.2 \\
         % & \ours{}         & \B{7.8} & 0.1 & \B{8.3} & 0.1 & \B{9.3} & 0.1 & \U{1.8} & 0.2 & \U{2.0} & 0.2 & \B{2.2} & 0.2 \\
         & \ours{} & \B{7.7} & 0.1 & \B{8.1} & 0.1 & \B{9.1} & 0.1 & \B{1.6} & 0.1 & \B{1.9} & 0.2 & \B{2.1} & 0.2 \\
\bottomrule
\end{tabular}
}
\caption{
\ac{vi} benchmark results. 
The $\pm$ symbol denotes a 95\% confidence interval.
}
\label{tab:version}
\end{table*}

\subsection{Training}\label{sec:baseline:training}
The architecture and \ac{cqt} inputs of \ours{} follow \clews{} exactly, except that we remove the frequency stride from the front-end and L2-normalize the embeddings.
At every iteration, we independently sample 220 unique segment cliques, with sampling probabilities weighted by the log-count of versions in the corresponding track clique.
From each segment clique, we sample an anchor segment from a uniformly chosen version, then sample an additional segment from a different version.
We use the triplet loss with the squared Euclidean distance and a margin of 0.2~\cite{schroff_facenet_2015}.
Every sample in the batch serves as an anchor, for which we mine the hardest negative.
We use the Adam optimizer and apply cosine annealing to the learning rate from $3\cdot10^{-4}$ to $3\cdot10^{-5}$ over 100,000 iterations.
Automatic mixed precision training with FP16 takes about 4 days on 4 NVIDIA L40S GPUs.

To improve robustness, we manipulate and degrade the training segments.
The ideal order should follow \figref{fig:signal}, which would slow down training.
Instead, we apply degradations to the waveform, extract the \ac{cqt}, and apply the manipulations to the \ac{cqt}.
Each operation is applied independently with 25\% probability, except time shifting, which is always applied: time shifting the segment by up to 1\,s; background noise, room reverberation, and microphone response (all drawn from the \nmfp{} train set), with noise mixed at an SNR drawn from $[0,\,12]\,\mathrm{dB}$; key transposition by $[0,\,12]$ semitones; time stretching, with a ratio drawn from $[0.55,\,1.80]$; and spectral masking along both the time and frequency axes, with each mask spanning a proportion of total frames or bins drawn from $[0,\,15]$\,\%.
% by extracting one additional \ac{cqt} octave, sampling a random integer shift and cropping the representation accordingly

\section{Results}\label{sec:results}

\tabref{tab:track} and \tabref{tab:version} report the results for \ac{ti} and \ac{vi}, respectively. In \tabref{tab:track}, the \ac{ti} models \nmfp{} and \peaknet{} show top performance on clean queries, whereas \clapp{} and the \ac{vi} models \ours{} and \clews{} follow.
\tabref{tab:version}, however, tells a different story: \ac{ti} models show essentially no \ac{vi} capability, while \ours{} and \clews{} achieve performance levels that can support retrieval.
\coverhunter{} fails on \ac{ti} and performs poorly on \ac{vi}, while \bytecover{} performs poorly across tasks.
We exclude \peaknet{} from the \ac{vi} evaluation given the benchmark cost.

The failure of \ac{ti} models on \ac{vi} is expected.
\ac{ti} embeddings are designed to identify a specific track, which results from a unique recording of a unique rendition of a unique composition.
The recording (equipment, environment, etc.) and rendering (performance, structure, etc.) introduce version-level variations, which \ac{vi} models suppress while preserving work identity.
Notably, \clapp{} fails on \ac{vi} despite performing competitively on clean and manipulated \ac{ti} queries, showing that generic audio representations do not necessarily encode work identity.
Recent work in musical influence attribution~\cite{barnett_exploring_2024} and data replication~\cite{batlle-roca_towards_2024} adopts \clapp{} as a proxy for musical similarity, yet our results show that \clapp{} fails at \ac{vi}.
% the most explicit case of musical relatedness.
% : two renditions of the same work.

In contrast, the \ac{ti} performance of \ours{} and \clews{} is noteworthy.
Traditionally, \ac{ti} has been targeted with fingerprinting methods that rely exclusively on acoustic features.
In \tabref{tab:track}, we demonstrate that the \ac{vi} models \ours{} and \clews{}, which suppress acoustic detail, can perform \ac{ti}.
First, on clean queries, \nmfp{} outperforms \ours{} by far on \hratk{1}, but the gap narrows considerably on \hratk{10}.
Note, however, that \nmfp{} yields 19 embeddings (1\,s context, 0.5\,s hop) for each 10\,s \ac{ti} query segment, while \ours{} and \clews{} yield a single one.
Section~\ref{sec:results:additional} analyzes this gap further.
Second, \ours{} and \clews{} are more robust to signal manipulation than \nmfp{}, which is not trained with manipulation.
Finally, against combined manipulation and degradation, \nmfp{} and \ours{} are the most robust models on \ac{ti}.
\ours{} outperforms \nmfp{} on the \discogsvi{} test set on both metrics, 
while on the \nmfp{} test set, \nmfp{} leads on \hratk{1} and \ours{} on \hratk{10}.
\nmfp{}'s \hratk{1} edge is consistent with domain match: its train and test set feature non-professional and non-music recordings.

Returning to \ac{vi} performance, \ours{} and \clews{} obtain the best results on clean queries and show strong robustness to manipulation.
Additionally, \ours{} demonstrates strong robustness to degradation.
Together, these results indicate that both models cluster versions of musical phrases, including their \mandeg{} versions.
Notably, \clews{} achieves this robustness without including degradation in its training pipeline, possibly leveraging the live versions and noisy recordings present in \discogsvi{}.
Nonetheless, \ours{}, which does include degradation, achieves substantially stronger robustness in \ac{vi}.
Moreover, on \discogsvi{}, \ours{} surpasses \clews{} even on clean queries, despite its supervision being derived from \clews{} embeddings, suggesting potential benefits of fully supervised training on segment cliques.

\subsection{Retrieval Mechanics Analysis}\label{sec:results:additional}
Our main results show that, despite its strong \ac{vi} performance, \ours{} underperforms \nmfp{} on \ac{ti} on clean queries.
Here, we examine this gap by performing exhaustive retrieval on the \discogsvi{} validation set.
Recall that the database tracks are segmented using a fixed hop duration for indexing (5\,s for \ours{}, 0.5\,s for \nmfp{}; Section~\ref{sec:eval:models}), and that the clean query segments in the \ac{ti} setup are randomly sampled from the same tracks (Section~\ref{sec:eval:query}).
This creates a random \textit{boundary misalignment} between the query and database segments, bounded by half the database hop duration (i.e., 2.5\,s for \ours{} and 0.25\,s for \nmfp{}).
Also recall that the database tracks are segmented with each model's training context duration, whereas the \ac{ti} query segments are 10\,s.
For \ours{}, variable-length processing embeds each query as a single vector, creating an \textit{information mismatch}: a query embedding summarizes 10\,s, whereas a database embedding, 20\,s.
This mismatch does not arise for \nmfp{} (1\,s context), whose query and database embeddings summarize the same duration.
We analyze these two factors, reporting track \hratk{1} (denoted simply as \hratk{1} in the main results) and introducing version \hratk{1}, where a hit is defined as a retrieved track being a version of the query.
Finally, we analyze the effect of reducing the segment duration on both \ac{ti} and \ac{vi}.

\vspace{0.1cm}\noindent
\textbf{Boundary misalignment:}
To measure performance under zero misalignment, we shift each query segment to the nearest point on the hop-duration grid.
We then perform retrieval with the randomly sampled queries and their grid-aligned variants, reporting the results in \tabref{tab:misalignment}.
Even under boundary misalignment, \nmfp{} achieves a near-perfect score on version \hratk{1}, and its track \hratk{1} recovers once misalignment is removed.
We attribute the remaining gap at zero misalignment for \nmfp{} to equivalent versions in the dataset (i.e., duplicates, remasters, extended mixes), which create ambiguity in the ground truth.
In contrast, for \ours{}, eliminating misalignment alone does not yield a perfect score on either metric.
We note that shortening the hop duration to reduce misalignment was computationally infeasible at the scale of test-set evaluation.

\begin{table}[t]
\captionsetup{skip=5pt, font=small}
\centering
\small
\setlength{\tabcolsep}{5.6pt}
\begin{tabular}{l S[table-format=1.2] cc}
\toprule
    Model & {Misalignment\,(s)} & Track \hratk{1} & Version \hratk{1} \\
\midrule
    \multirow{2}{*}{\ours{}}
     & 0.00 & 95.0 & 98.5 \\
     & 2.50 & 91.9 & 98.2 \\
\midrule
    \multirow{2}{*}{\nmfp{}}
     & 0.00 & \B{99.7} & \B{99.8} \\
     & 0.25 & 96.7 & 99.8 \\
\bottomrule
\end{tabular}
\caption{
Effect of boundary misalignment between query and database segments on \ac{ti} performance.
Database segments match each model's training context duration; information mismatch is present for \ours{}.
Misalignment values are upper bounds.
}
\label{tab:misalignment}
\end{table}

\vspace{0.1cm}\noindent
\textbf{Information mismatch:}
To also eliminate the information mismatch for \ours{}, we re-extract the database embeddings using 10\,s segments (the query duration) and a 5\,s hop.
We then evaluate on both the original queries and their grid-aligned variants, reporting the results in \tabref{tab:mismatch}.
Under boundary misalignment, track \hratk{1} is much lower than version \hratk{1}, i.e., versions of the reference track rank closer to the query than the reference track itself does.
Once misalignment is also removed, \ours{} obtains near-perfect scores, up to the ambiguity introduced by equivalent versions.
Given \tabref{tab:misalignment} and \tabref{tab:mismatch}, we attribute the \ac{ti} performance gap of \ours{} in \tabref{tab:track} to retrieval constraints rather than a limitation of the embedding space.

\begin{table}[t]
\captionsetup{skip=5pt, font=small}
\centering
\small
\setlength{\tabcolsep}{11pt}
\begin{tabular}{ccc}
\toprule
    Misalignment\,(s) & Track \hratk{1} & Version \hratk{1} \\
\midrule
    0.0 & \B{99.9} & \B{100.0} \\
    2.5 & 92.7 & ~~99.6 \\
\bottomrule
\end{tabular}
\caption{
Effect of boundary misalignment on \ac{ti} performance for \ours{} with 10\,s database segments, which eliminates the information mismatch.
Misalignment values are upper bounds.
}
\label{tab:mismatch}
\end{table}

\vspace{0.1cm}\noindent
\textbf{Segment duration:}
Having seen that \ours{} benefits on \ac{ti} from database segments matching the query duration, we now extract the query and database embeddings for both tasks with a 10\,s segment duration, keeping a 5\,s hop.
We also reproduce the original setting (20\,s segments, 5\,s hop) on this set for comparison, where each 10\,s \ac{ti} query is embedded as a single vector.
\tabref{tab:segment} reveals a trade-off for \ours{}: shorter segments improve \ac{ti} performance but substantially degrade \ac{vi} performance.

\begin{table}[t]
\captionsetup{skip=5pt, font=small}
\centering
\small
\setlength{\tabcolsep}{5.5pt}
\begin{tabular}{ccccc}
\toprule
    \multirow{2}[+1]{*}{Dur.\,(s)} & \multicolumn{2}{c}{\ac{ti}} & \multicolumn{2}{c}{\ac{vi}} \\
\cmidrule(lr){2-3}\cmidrule(lr){4-5}
     & Clean & Manip. \& Deg. & Clean & Manip. \& Deg. \\
\midrule
    10 & \B{92.7} & \B{74.6} & 0.529 & 0.390 \\
    20 & 91.9 & 69.2 & \B{0.780} & \B{0.713} \\
\bottomrule
\end{tabular}
\caption{
Effect of segment duration on the \ac{ti} and \ac{vi} performance of \ours{}, applied to the database tracks and to \ac{vi} queries; \ac{ti} queries are 10\,s in both settings.
Track \hratk{1} and \mapat{10\,k} are reported for \ac{ti} and \ac{vi}, respectively.
}
\label{tab:segment}
\end{table}

\subsection{Data Construction Ablation}\label{sec:results:ablation}

The training data construction pipeline has multiple steps (Section~\ref{sec:baseline:data}).
In \tabref{tab:data}, we analyze the effect of training on the data produced by each stage.
$\mathcal{D}_1$ contains all segment pairs with distance below a static threshold of 1.0, across all version pairs.
$\mathcal{D}_2$ instead applies a dynamic threshold per version pair, computed as the 5\textsuperscript{th} percentile of its distance distribution.
$\mathcal{D}_3$ is the subset of $\mathcal{D}_2$ restricted to the local minima located by the flood-fill algorithm.
$\mathcal{D}_4$ groups the local minima of $\mathcal{D}_3$ across versions into segment cliques based on temporal overlap.
On each dataset, we train for 40,000 iterations using a single GPU and run the validation routine.
The results reported so far for \ours{} were obtained with hard negative mining, which is notoriously difficult to stabilize~\cite{schroff_facenet_2015}.
Since the earlier stages of the data pipeline produce noisier data, we use semi-hard negative mining to provide a stable basis for comparison.

\begin{table}[t]
\captionsetup{skip=5pt, font=small}
\centering
\small
\setlength{\tabcolsep}{3.8pt}
\begin{tabular}{cccccc}
\toprule
    \multirow{2}[+1]{*}{Data} & \multirow{2}[+1]{*}{N.M.} & \multicolumn{2}{c}{\ac{ti}} & \multicolumn{2}{c}{\ac{vi}} \\
    \cmidrule(lr){3-4}\cmidrule(lr){5-6}
     &  & Clean & Manip. \& Deg. & Clean & Manip. \& Deg. \\
\midrule
    $\mathcal{D}_1$ & SH & $86.5$ & $40.9$ & $0.551$ & $0.359$ \\
    $\mathcal{D}_2$ & SH & $86.1$ & $45.7$ & $0.611$ & $0.469$ \\
    $\mathcal{D}_3$ & SH & $81.3$ & $46.7$ & $0.697$ & $0.552$ \\
    $\mathcal{D}_4$ & SH & $84.5$ & $51.6$ & $0.686$ & $0.550$ \\
    $\mathcal{D}_4$ & H  & $\B{90.3}$ & $\B{64.6}$ & $\B{0.738}$ & $\B{0.645}$ \\
\bottomrule
\end{tabular}
\caption{
Effect of training data construction on \ours{}.
\hratk{1} and \mapat{10\,k} metrics are reported.
N.M.: Negative mining strategy; SH: semi-hard, H: hard.
% All models are trained for 40,000 iterations to enable direct comparison.
}
\label{tab:data}
\end{table}

\tabref{tab:data} shows that the dynamic threshold ($\mathcal{D}_2$) substantially improves \ac{vi} performance over the static threshold ($\mathcal{D}_1$), and restricting $\mathcal{D}_2$ to local minima ($\mathcal{D}_3$) yields further gains.
Grouping local minima ($\mathcal{D}_3$) into segment cliques ($\mathcal{D}_4$) shows no clear benefit under semi-hard negative mining.
However, hard negative mining on $\mathcal{D}_4$ (used in the main experiments) achieves the best results.
Interestingly, hard negative mining on $\mathcal{D}_3$ converges, but inference is unstable, and we cannot characterize the root cause.
Overall, using the data effectively is not straightforward.

\section{Conclusion}\label{sec:conclusion}
Our unified benchmark evaluates \ac{ti} and \ac{vi} performance under both signal manipulation and audio degradation.
In this experimental setup, in which 10\,s of query audio is available for \ac{ti}, we conclude that the \ac{vi} models \clews{} and \ours{} can perform \ac{ti}.
Subsuming \ac{ti}, however, requires overcoming the boundary misalignment and information mismatch factors, which, for \ours{}, we showed to be retrieval constraints rather than a limitation of the embedding space.
The \ac{ti} model \nmfp{} avoids both factors through its short hop and context durations, at the expense of a larger index (10$\times$ more embeddings) and more searches per unknown recording (19$\times$), although its smaller embedding dimensionality (8$\times$) reduces the number of operations per search.
Future work could search for an optimal training context duration for a unified model, since matching the database segment duration to the \ac{ti} query duration degrades the \ac{vi} performance of \ours{}.
The performance of such a model for query durations shorter than 10\,s remains an open question: if \ac{vi} indeed imposes a lower bound on the segment duration, shorter queries would necessarily incur information mismatch.
Meanwhile, a stronger \ac{ti} model that is robust to signal manipulation may make the benchmark harder, although such a model is unlikely to show meaningful \ac{vi} performance.

\clearpage

\section{Acknowledgments}
We thank Emilio Molina Martínez and Pablo Zinemanas from the BMAT Music Innovators team for the helpful discussions throughout this project, and Davide Scaini for his valuable input on designing the signal manipulation and audio degradation pipeline.
R.~Oguz Araz is partially supported by the pre-doctoral grant AGAUR-FI Joan Oró (2024 FI-3 00065); the Cátedra IA y Música project (TSI-100929-2023-1), funded by the Secretaría de Estado de Digitalización e Inteligencia Artificial and the European Union's NextGenerationEU funds and by BMAT Music Innovators; and the TROBA project (ACE014/20/000051), funded by ACCIÓ - Nuclis d’R+D 2024.

\bibliography{Unified}

\end{document}